# ORIGINAL ARTICLE

# Application of Graph Based Features in Computer Aided Diagnosis for Histopathological Image Classification of Gastric Cancer

Haiqing Zhang,[1] Chen Li,[1,*] Shiliang Ai,[1] Haoyuan Chen,[1] Yuchao Zheng,[1] Yixin Li,[1] Xiaoyan Li,[2] Hongzan Sun,[2] Xinyu Huang,[3] Marcin Grzegorzek[3]

[1]*Microscopic Image and Medical Image Analysis Group, College of Medicine and Biological Information Engineering, Northeastern University, Shenyang, China*
[2]*China Medical University, Shenyang, China*
[3]*Institute of Medical Informatics, University of Luebeck, Luebeck, Germany*

[*]Correspondence should be addressed to Chen Li; lichen201096@hotmail.com

## Abstract

The gold standard for gastric cancer detection is gastric histopathological image analysis, but there are certain drawbacks in the existing histopathological detection and diagnosis. In this paper, based on the study of computer aided diagnosis system, graph based features are applied to gastric cancer histopathology microscopic image analysis, and a classifier is used to classify gastric cancer cells from benign cells. Firstly, image segmentation is performed, and after finding the region, cell nuclei are extracted using the *k*-means method, the minimum spanning tree (MST) is drawn, and graph based features of the MST are extracted. The graph based features are then put into the classifier for classification. In this study, different segmentation methods are compared in the tissue segmentation stage, among which are Level-Set, Otsu thresholding, watershed, SegNet, U-Net and Trans-U-Net segmentation; Graph based features, Red, Green, Blue features, Grey-Level Co-occurrence Matrix features, Histograms of Oriented Gradient features and Local Binary Patterns features are compared in the feature extraction stage; Radial Basis Function (RBF) Support Vector Machine (SVM), Linear SVM, Artificial Neural Network, Random Forests, *k*-NearestNeighbor, VGG16, and Inception-V3 are compared in the classifier stage. It is found that using U-Net to segment tissue areas, then extracting graph based features, and finally using RBF SVM classifier gives the optimal results with 94.29%.

## INTRODUCTION

**Background**

Cancer is a disease caused by uncontrolled growth of cells [1], while gastric cancer is a group of abnormal cells that gather in the stomach to form tumor. Comprehensive data from recent years show that the incidence and mortality rate of gastric cancer is the third highest among women and the second highest among men.

Effective diagnosis of gastric cancer relies on the examination of hematoxylin and eosin (H&E) stained tissue sections under a microscope by pathologists. Microscopic examination of tissue sections is tedious and time-consuming, with screening procedures usually taking 5-10 minutes, making it difficult for pathologists to analyse more than 70 samples a day [2]. And the potential for incorrect diagnosis is high. Therefore, histopathological diagnosis of gastric cancer is important [3, 4].

Computer aided diagnosis (CAD), the basic concept of it is to use computed judgments as objective opinions to help pathologists make a diagnosis [5-8]. The goal of CAD is to improve the quality and efficiency of histopathological images. And through increasing the accuracy and consistency of image diagnosis, it can reduce image reading time [9-14]. In the last few decades, a lot of research has been done on the development of CAD system that can help physicians track cancers [15-18]. Meanwhile, cancerous gastric histopathological images cells proliferate indefinitely [19]. It causes cancer cells to become denser, and the graph made of plasmas with nuclei is more compact than normal cells. Therefore, the study of graph theory is applied to classify histopathological images with better results.

In this paper, a method of classifying histopathological images of gastric cancer using graph based features is proposed. The workflow is shown in Figure 1.

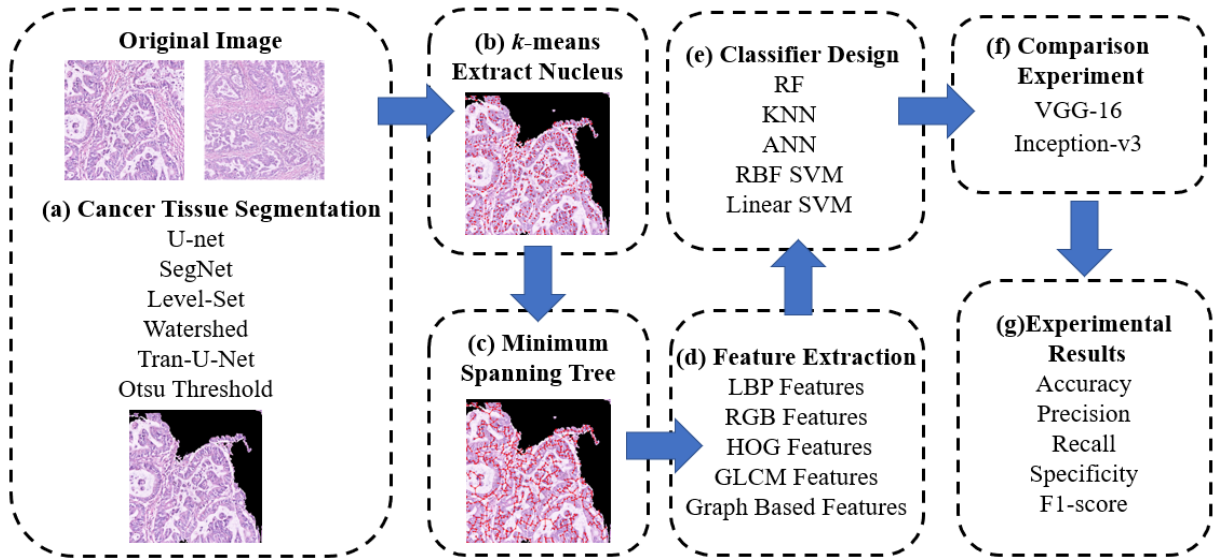

Figure 1: Workflow of the proposed method.

**Research Content**

The method proposed in this paper consists of seven main parts: (a) Six different image segmentation methods are compared to obtain the optimal one. (b) Cell nuclei are extracted using *k*-means clustering method. (c) The center of mass of the cell nuclei is used as the point to form the graph structure and extract the graph based features using the minimum spanning tree (MST) algorithm. (d) Graph based features, Red, Green, Blue (RGB) features, Gray-level Co-occurrence Matrix (GLCM) features, Local Binary Pattern (LBP) features, and Histogram of Oriented Gradient (HOG) features are extracted after segmentation for a comparison. (e) Based on the previous work, the extracted feature vectors are put into different classifiers: Radial Basis Function (RBF) Support Vector Machine (SVM), Linear SVM, Artificial Neural Network (ANN), Random Forests (RF) and *k*-Nearest Neighbor (KNN). (f) Two deep learning comparison experiments are designed. (g) Experimental results are obtained by calculating Accuracy, Precision, Recall, Specificity, and F1-score.

The main contributions of this paper are as follows:

- A new framework is designed to introduce graph based features image classification method to the field of histopathological image analysis of gastric cancer.
- A large number of comparative experiments are done to demonstrate the effectiveness of our method.
- This method for gastric cancer proposed in this study achieves good results, with a final classification accuracy of 94.29%.

## RELATED WORKS

**The Development of Graph Theory on CAD**

The application of graph theory in CAD can effectively use the topological structure of histopathological images to analyse the information in histopathological images by analysing the structure of the graph. And the length and size of each edge and corner in the graph can be used to represent the spatial ordering of different tissues, which can intuitively reflect the content of histopathological images. These data can be better used as the basis for the judgment of pathologists. Therefore, the application of graph theory techniques to CAD analysis of histopathological images has become popular.

Graph theory is applied to extract topological and relational information from collagenz frameworks through the integration of deep learning with graph theory [20]. The results are consistent with the expected pathogenicity induced by collagen deposition, demonstrating the potential for clinical application in analysing various reticular structures in whole-slide histological images.

Computer image-processing-based Voronoi diagrams, Gabriel diagrams and MST are used to represent the spatial arrangement of blood vessels as a way to quantitatively analyse microvessels [21]. Derived features of graph structure are extracted using syntactic structure analysis of graph structure-based derived features. The most discriminative features are found using a *k*-nearest neighbour classifier.

A large number of colour, texture and morphological features are extracted from stained histopathological images of cervical cancer [22]. And it extracts 29 features such as edge length and area from three graph structures, after which the nuclei are classified using linear discriminant analysis.

Instead of cell nuclei, skeletal nodes are used in histopathological images of cervical cancer, and the work constructs graph structures using MST, extracts several feature values such as the edge and corner, and clusters them using *k*-means [23]. In addition, four graph theoretic methods are added as a comparison in the step of constructing the graph structure to select the optimal graph theoretic method. The deep learning network structure with gradient direction histogram is used to compare with the selected optimal graph theoretic methods, and the optimal method is obtained by the evaluation of doctors.

**The Development of CAD in Gastric Cancer**

Deep learning is used in many medical image processing tasks, for example, deep learning is used to identify COVID-19 samples in chest x-ray images [24]. The continuous progress of deep learning algorithms has led to the rapid development of CAD technology in gastric cancer. Currently, the deep learning methods used in the field of gastric cancer mainly include the following aspects: image pre-processing, image segmentation, feature extraction, and image classification methods.

In histopathological image pre-processing of gastric cancer, the work proposes an image classification model, which can alleviate the bad annotation training set [25]. Through fine-tuning the neural network in two stages and introducing a new intermediate dataset, the performance of the network in image classification is improved. The work sets up an image-denoising network based on CNN and optimizes the denoising network by using the advantages of complex numerical operations to increase the tightness of convolution [26].

In the image segmentation stage, a new polyp segmentation method based on multidepth codec network combination is proposed [27]. The network can extract the features of different effective receptive fields and multi-size the image to represent multi-level information. It can also extract effective information features from the missing pixels in the training phase. A radiology-based deep supervised U-Net is developed for the segmentation of prostate and prostatic lesions [28]. These methods can also be applied to histopathological images of gastric cancer. A highly effective hybrid model for filter bubble partitioning is proposed [29].

In the stage of feature extraction, the HOG and LBP features are extracted in gastric cancer histopathological images [30]. By comparing, the LBP feature is superior to the HOG feature. A new unsupervised feature selection method, which calculates the dependencies between features and avoids selecting redundant features is developed [31]. It can also be used in the histopathological images of gastric cancer to improve the operation efficiency.

In terms of classifiers, this paper compares RF and ANN. Through extensive experimental comparisons, the artificial neural network classifier outperforms the RF classifier. The standard Inception-V3 network framework is used [32]. The parameters are reduced by changing the depth multiplier. After several iterations, the model with the lowest validation error is finally selected as the final model.

## METHOD

**U-Net Based Image Segmentation**

U-Net is an FCN-based semantic segmentation network originally applied to medical cell microscopy image segmentation tasks [33-35]. The end-to-end structure of this network can efficiently recover the information loss at the shallow level due to pooling operations. In addition, the training strategy of U-Net network uses a data expansion strategy that efficiently makes full use of the limited labelled training data for de-training. U-Net contains two parts, the first part is the left half of U-shaped structure for feature extraction. The second part is the right half of the U-shape, which is the up-sampling part. A copy and crop jump connection layer is used before fusion to ensure that more features are fused in the final recovered feature map, which also allows the fusion of features of different sizes, thus allowing multi-scale prediction.

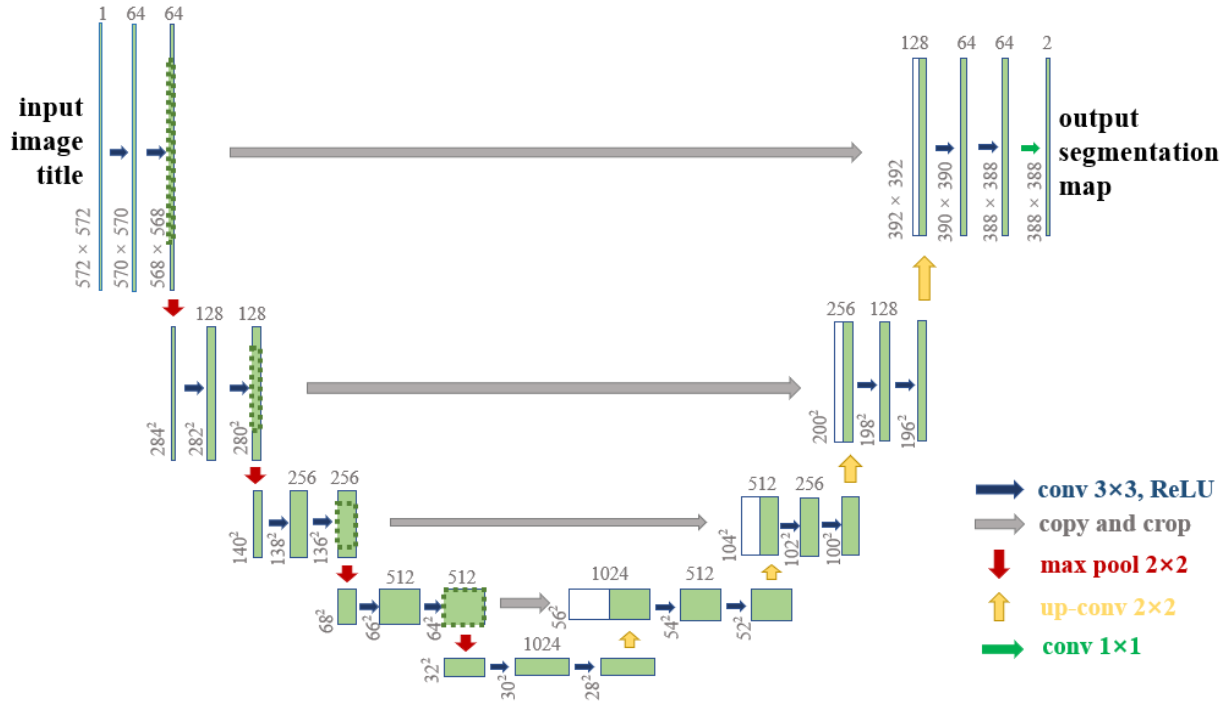

Figure 2: U-Net structure.

Figure 2 shows U-Net structure used in this paper. The network consists of a down-sampling (left side) and an up-sampling (right side) path. In the down-sampling path, a convolution operation using ReLU for activation is first followed by a max pooling operation. After that, the size of the feature map generated by the convolution operation is reduced to half of the original size. This set of operations is repeated four times in the down-sampling path. In the up-sampling path, each step contains three main operations, which are the up-convolution operation, the copy and stitch operation, and the convolution operation with ReLU for activation. These three operations are repeated a total of four times in the up-sampling path. The final segmentation result is generated by a $2 \times 2$ sized convolution kernel (activated using Sigmoid).

**Graph Theory and Graph Based Features**

Graph theory takes graphs as the object of study [36]. A graph is a figure consisting of a number of given nodes and an edge connecting the two nodes. Such graphs are usually used to describe a particular relationship between certain things, with the points representing the things and the edges connecting the two nodes indicating that the corresponding two things have this relationship. A graph usually contains nodes, edges, paths, loops, and weights. An example of a graph is shown in Figure 3.

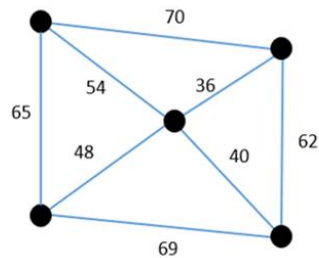

Figure 3: An example of a graph with five points and eight weighted edges.

Graph theory is a graph structure which can record the topological structure in an image by computing the features of the graph structure. There are various ways of graph composition: MST [37], Delaunay triangulation [38], Voronoi Diagram [39], etc. Our previous work finds that the graph formation characteristics of MST are better in comparison with cervical cancer histopathological images [23,40]. A comparative analysis reveals that the topological information carried by using the minimum spanning tree graph structure in histopathology is the most complete [40]. So the MST graph based features obtains the optimal result. Therefore, in this work, the MST is chosen as the graph formation method for the graph structure.

This paper proposes a method for image analysis of histopathology of gastric cancer using graph based features. While observing the experimental data, the cancerous tissues in the histopathological images of gastric cancer are significantly different from the normal images in terms of topological structure. Therefore, we intend to design a method to classify the gastric cancer pathology using topological structure. A large amount of literature shows that the topological structure information in the graph can be gotten by the method of graph theory, and the MST algorithm is chosen as the graph-forming method.

In this work, the information of edges and angles is obtained on the MST. The reason why extracting the information of edge lengths and angles is that the graph is composed of nodes and edges and more than 2-3 nodes constitute a corner. Edges and angles are the most basic elements to characterize the graph structure. Edges represent the degree of dispersion between two nodes, and angles represent structural complexity [41]. The information of edges is the edge length of the MST, and the information of angles is to calculate the angle between every two adjacent edges. The MST contains all nodes with the smallest sum of weights in the original graph, which is the least connected subgraph of the Delaunay triangulation. At the level of the sum of edge weights, the value of MST is less than or equal to the sum of the weights of all other spanning trees. As shown in Figure 4, there are five points ABCDE, and ∠ABC, ∠ABD, ∠CBD and ∠BDE can be calculated. Mean, variance, kurtosis and skewness are found for all edges and angles of each tree, and eight feature values can be output for each histopathological image.

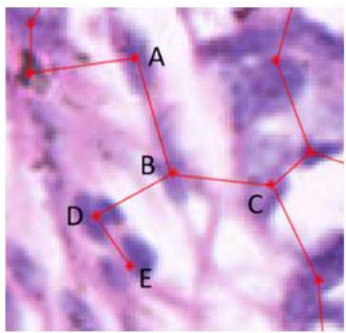

Figure 4: Example Figure of a MST composed of cell nuclei.

**Classification Methods**

In this paper, we compare the performance of different kinds of classifiers, including RBF SVM, Linear SVM, ANN, KNN, RF, and two classification models in deep learning: the VGG-16 and Inception-V3. SVM is a binary classifier that uses linearity for classification [42], and it produces different classification results depending on the kernel function. The Linear kernel function, on the other hand, has the advantage of having fewer parameters and fast evolution on linearly separable data. ANN is a model that simulates the information transfer mechanism of neurons [43], which has the advantage of high accuracy and parallel distribution processing capability but requires many initial parameters and a long training time. KNN is a machine learning a simple classification learning algorithm [44], which is suitable for multi-label problems, but the accuracy suffers when the number of samples of classes is unbalanced. RF random forest is an integrated learning method in machine learning [45]. RF is simple, easy to implement, and has low computational overhead, but is prone to overfitting.

VGG-16 uses small-sized convolutional kernels to reduce the parameters, and the network structure is regular and suitable for parallel acceleration [46]. The main idea of Inception architecture is to find out how to approximate the optimal local sparse knots with dense components. This network model proposed by Inception-V3 gives more accurate feature information when dealing with a larger number of features and features with high feature diversity, and also reduces the computational effort [47].

The SVM classifier with RBF kernel function is chosen for the experiments. The main reasons are: SVM classification is effective, the most fundamental verdict in SVM classification judgments is provided by support vectors, the complexity of the computation is not affected by the number of samples, but mainly by the number of support vectors, so the structure has small storage space, and the algorithm is robust. Also, SVM is a small sample learning method that does not involve concepts such as probability measures, simplifying the usual problems such as classification and regression. In terms of kernel functions, the RBF kernel function is more advantageous on linearly indistinguishable data, and classification is more accurate.

## Experiments and Analysis

**Image Dataset and Experimental Setup for Gastric Cancer**

The 2017 BOT competition provides 700 histopathological images with 2048×2048 resolution. Among them, 560 are histopathological images of gastric cancer and the rest are histopathological images of normal stomach. The training set of U-Net segmentation network contains 300 images randomly selected among 560 histopathological images of gastric cancer. 120 are the validation set and the remaining 140 are used as the test set. Then, experimental operations such as feature extraction are performed on the basis of these 140 histopathological images of gastric cancer and 140 histopathological images of normal stomach separately. In the classifier comparison section, data augmentation is performed on the existing data to improve the performance of the classifier.

During the training, only abnormal images have GT images, and normal images do not have GT images. But in the later test, all images are subjected to image segmentation and graph based features are calculated.

**Analysis of Image Segmentation Results**

In the image segmentation stage, the same two images are segmented in this paper using six segmentation methods respectively. In this paper, the parameters of U-Net are set such that two sets of convolution operations of size $3 \times 3$, followed by Max Pooling operations of size $2 \times 2$. The size of the up-convolution operation is $3 \times 3$, and the third operation is two $1 \times 1$ convolution operations. These three operations are repeated a total of four times in the up-sampling path. The final segmentation result is produced by a convolution kernel of size 2 × 2. The results of the different segmentation methods are shown in the Figure 5.

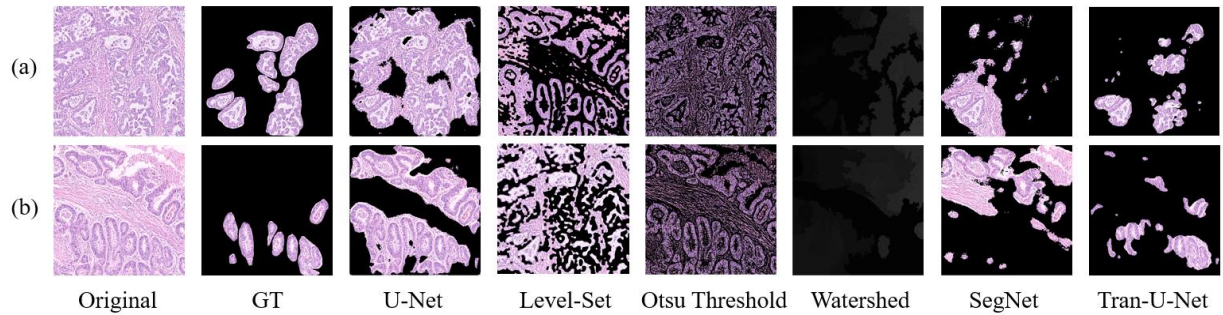

Figure 5: Comparison of six segmentation methods on the test set for 2 typical examples (a) and (b).

In Figure 5, it can be seen that the Level-Set segmentation method is segmented according to the edges of the image, which cannot distinguish the structure of different tissues in the pathological images and watershed segmentation only separates the whole image, but there is no gastric image segmentation method that is based on tissue structure. And the information collected is both good and bad, which cannot highlight the graph based features of the cancer region. The foreground retained by the Otsu thresholding segmentation method includes not only the effective tissue but also the intercellular matrix, which seriously affects the quality of the minimum generated tree graph structure. And the foreground retained by the Otsu thresholding segmentation method includes not only the effective tissue but also the intercellular matrix, which seriously affects the quality of the minimum generated tree graph structure. It demonstrates that U-Net method we used is able to segment the cancer region better, with smoother edges and clear subject regions, and retains less noise compared to other methods. In addition to the visual comparison, we also make evaluation metrics, the details of which are shown in Table 1.

Table 1: The accuracy of the image segmentation method. The first column shows the sequence number. The second column shows the six image segmentation methods used in this paper. The third to ninth columns show common metrics for evaluating segmentation, dice coefficient, intersection over union (IoU), precision, recall, specificity, RVD and pixel accuracy.

| No. | Segmentation Method | Dice | IoU | Precision | Recall | Specificity | RVD | Accuracy |
|---|---|---|---|---|---|---|---|---|
| 1 | Level-Set [48] | 0.2845 | 0.1920 | 0.2949 | 0.5202 | 0.7284 | 2.9296 | 0.6982 |
| 2 | Otsu Threshold [49] | 0.2534 | 0.1505 | 0.2159 | 0.4277 | 0.7082 | 2.8859 | 0.6598 |
| 3 | Watershed [50] | 0.2613 | 0.1585 | 0.2932 | 0.3541 | 0.7942 | 1.9434 | 0.7205 |
| 4 | SegNet [51] | 0.2008 | 0.1304 | 0.3885 | 0.3171 | 0.8412 | 2.0662 | 0.7531 |
| 5 | U-Net [33] | **0.4557** | **0.3191** | **0.4004** | **0.6896** | 0.7795 | **1.6736** | 0.7684 |
| 6 | Trans-U-Net [52] | 0.3295 | 0.2047 | 0.3815 | 0.3584 | **0.9051** | 2.6959 | **0.8125** |

Medical images are characterized by simpler semantics, fixed structure and less data volume. U-Net segmentation uses U-shaped structure and skip-connection to achieve more excellent performance, which can perfectly solve these problems and is very outstanding in the field of medical image segmentation. From the Table 1, except for specificity and accuracy, U-Net performs better than other methods in other indicators and lighter than Trans-U-Net.

**Analysis of *k*-means Algorithm**

At this stage of the paper, the pixel grayscale values of the images are clustered, and the *k*-means algorithm with *k*=3 is used as a benchmark for comparison. As shown in Figure 6, we can observe that when *k*=3, the nuclei in the histopathological images of gastric cancer are better expressed, and basically all the nuclei on the tissue are labelled. When *k*=4, a part of the stained gastric cancer tissue region is not labelled due to further clustering of the gray scale values. When *k*=5, this becomes more obvious and the nuclei information is severely lost. Therefore, this paper selects the *k*-means clustering algorithm with a better effect of *k*=3 to extract the cell nuclei in the extraction stage.

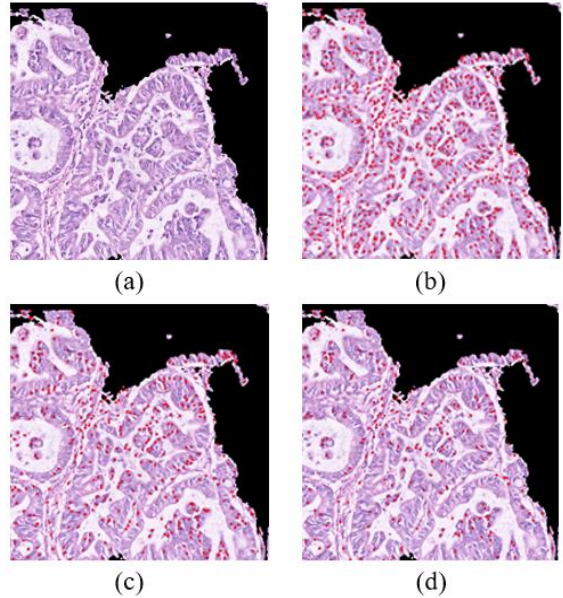

Figure 6: Cell nuclei results with different *k*.
(a) The upper right quarter area of the image, (b) The image of the nucleus with *k*=3,
(c) The image of the nucleus with *k*=4    , (d) The image of the nucleus with *k*=5.

**Analysis of Feature Extraction Methods**

After segmentation, the graph based features are used for feature extraction: *k*-means is used to extract the cell nucleus and the MST algorithm is used to draw the graph, which is shown in the Figure 7.

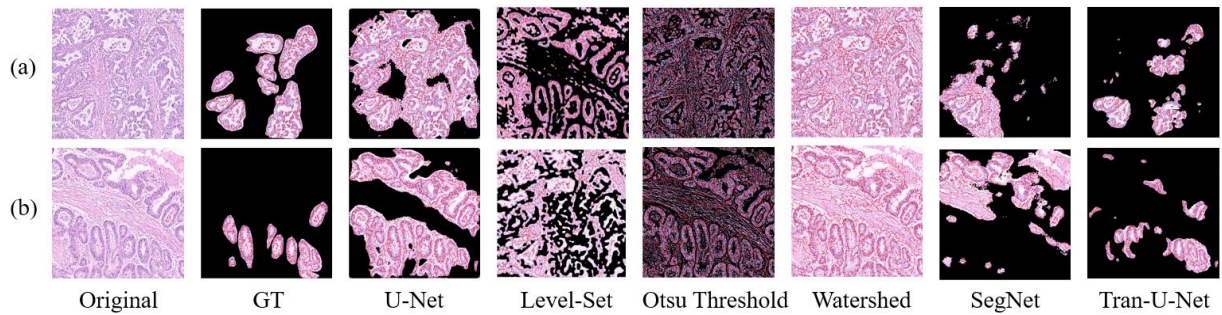

Figure 7: Comparison of MST graph structures of six segmentation methods on the test set for 2 typical examples (a) and (b).

By comparing the MST graph structures after the segmentation, the MST graph structure extracted from U-Net segmented images could represent the real topological information structure of gastric cancer tissues more accurately. But other segmentation methods have poor quality of their MST graph structures due to their own drawbacks.

Then, five feature extraction methods are compared in this paper, as shown in Table 2. Firstly, U-Net segmentation method is used for segmentation. After that, the graph based features (MST), RGB features, HOG features, GLCM features and LBP features are extracted and compared for the image.

Table 2: Accuracy of image segmentation method. The first column shows the serial number. The second column shows the segmentation using U-net. The third column shows the five feature extraction methods used in this paper. The fourth column shows the classification with RBF SVM classifier. The fifth column shows the accuracy rates corresponding to the results of different feature extraction methods in the experiments.

| No. | Segmentation Method | Feature Extraction | Classifier | Accuracy |
| --- | --- | --- | --- | --- |
| 1 |  | Graph Based Features (MST) |  | 94.29% |
| 2 |  | RGB Features |  | 69.29% |
| 3 | U-Net Segmentation | HOG Features | RBF SVM | 55.00% |
| 4 |  | GLCM Features |  | 53.57% |
| 5 |  | LBP Features |  | 65.71% |

Then, the RBF SVM classifier is selected for classification in the third step. Finally, by calculating the classification accuracy, it can be seen that the graph based features have obvious advantages in the feature extraction stage and the classification accuracy reaches 94.29%.

*Analysis of Graph Based Features*

This study uses two features of the MST that can represent the topology of a graph: the edge lengths and angles of the MST. Based on this, eight statistical features are extended, including the mean, variance, skewness and kurtosis of the edge length and the mean, variance, skewness and kurtosis of the angle.

As shown in Figure 8, the first column shows the characteristic statistics of the edge length, which are mean, variance, skewness and kurtosis. The second column represents the characteristic statistics of the angle, including mean, variance, skewness and kurtosis. The horizontal coordinates of the statistical plot indicate the number of images in the experiment, 140 in total. The first 70 are normal gastric histopathological images while the last 70 are gastric cancer histopathological images. The vertical coordinates of the statistical plot indicate the statistical values.

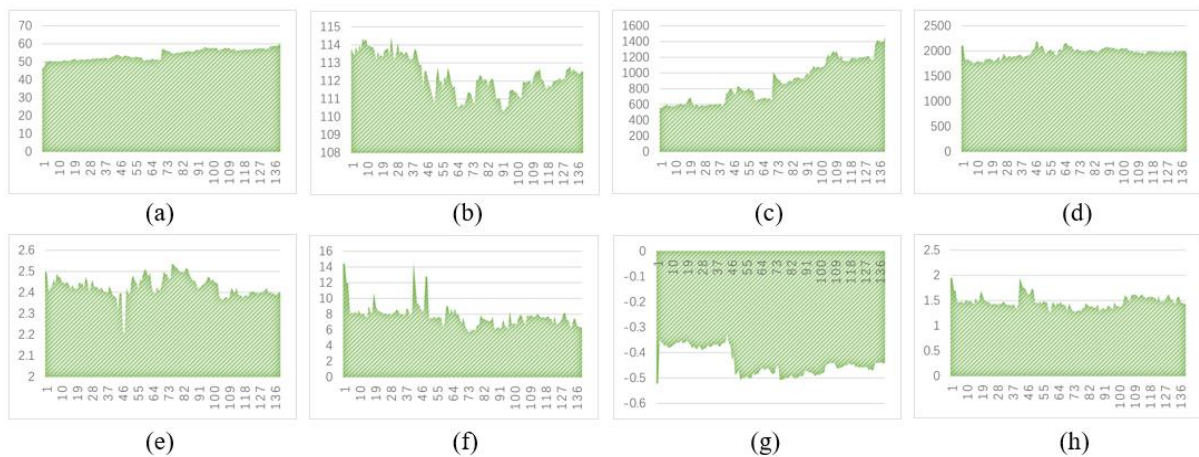

Figure 8: Statistical features of graph structure. The horizontal coordinates of the statistical plot indicate the number of images in the experiment. The vertical coordinates of the statistical plot indicate the statistical values.
(a) Statistical plot of the mean characteristics of the edge length, (b) Statistical plot of mean characteristics of the angle,
(c) Statistical plot of variance characteristics of the edge length , (d) Statistical plot of variance characteristics of the angle,
(e) Statistical plot of skewness characteristics of the edge length, (f) Statistical plot of skewness characteristics of the angle,
(g) Statistical plot of kurtosis characteristics of the edge length , (h) Statistical plot of kurtosis characteristics of the angle.

Meanwhile, it can be seen that the mean values of edge lengths of normal and gastric cancer histopathological images are more stable, indicating that the size of tissue structure is more accurately described, and the size of structure of each histopathological image does not differ greatly. In terms of variance, the variance of edge length of normal images is significantly smaller than that of gastric cancer images, indicating that the edge length structure of normal images is more similar, while the edge length structure of gastric cancer images is of different lengths and full of irregular shapes of cancer. And the angle of normal images is similar to that of gastric cancer images, and their angle structures are similar. In terms of skewness, the edge lengths of the normal and gastric cancer images are also relatively close. It indicates that they have a similar degree of asymmetry relative to the mean, while the angles are significantly different and they have a greater degree of asymmetry relative to the mean. In terms of kurtosis, the edge lengths and angles of normal and gastric cancer images are similar, indicating that the steepness of their distribution patterns are similar.

Collectively, it can be seen that the topological information of histopathological images can be extracted more completely by using graph based features (MST). The classification accuracy of this method is the highest among all the feature extraction methods, reaching 94.29%. And it fully illustrates the high performance of the graph based feature extraction method on the histopathological images of gastric cancer.

*Analysis of RGB Features*

In this study, in the RBG features extraction method, the histogram statistics of R channel, G channel and B channel of each image are performed. As shown in Figure 9, the horizontal coordinate of this histogram is the pixel value (the interval is from 0 to 255) and the vertical coordinate is the number of pixels for each pixel value. In the statistics of each channel, the background (the pixel value is 0) in U-Net segmentation images is removed, and this region is the part that is segmented off during U-Net segmentation process and cannot be counted into the RGB features as feature statistics.

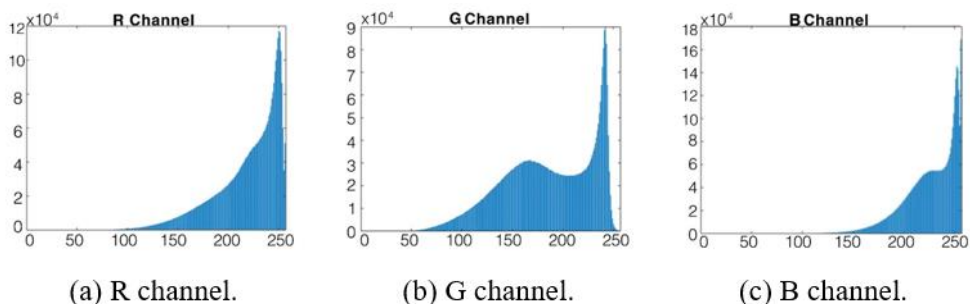

(a) R channel.    (b) G channel.    (c) B channel.

Figure 9: RGB features histogram. The horizontal coordinate of this histogram is the pixel value (the interval is from 0 to 255) and the vertical coordinate is the number of pixels for each pixel value.

*Analysis of HOG Features*

In this study, feature vectors of dimension 2340900 are extracted for each U-Net segmentation graph in the HOG feature extraction stage and then put into RBF SVM for classification. The HOG features use grayscale gradients to describe the local shape of the object, which marks the gradient oritation of the graph on the original graph. The advantage of HOG features is that they have better geometric invariance with optical undistorted due to the histogram of gradient oritations drawn in a small region. The disadvantage is that the noise immunity is poor. HOG features are better in detecting human body according to their characteristics, but the effect of feature extraction for histopathological images is not satisfactory. The classification accuracy of feature extraction is only 55%.

Considering that the feature dimensions extracted by different classifiers are different, the effect of different classifiers is also compared in HOG feature extraction, which is shown in the Table 3.

Table 3: The results of each metric obtained by classifying HOG features using different classifiers. The main evaluation metrics consist of the following five components: Accuracy (Acc), Precision (Pre), Recall (Rec), Specificity (Spe) and F1-Score. And they are different in abnormal and normal images.

| Methods | Acc | Abnormal | | | | Normal | | | |
|---|---|---|---|---|---|---|---|---|---|
| | | Pre | Rec | Spe | F1 | Pre | Rec | Spe | F1 |
| Linear SVM | 53.28 | 44.82 | 80.14 | 35.79 | 57.49 | 73.47 | 35.79 | 80.14 | 48.13 |
| non-linear SVM | 60.58 | Null | 0.00 | 100.00 | 0.00 | 60.58 | 100.00 | 0.00 | 75.45 |
| RF | 60.85 | 50.33 | 53.01 | 65.95 | 51.63 | 68.32 | 65.95 | 53.01 | 67.11 |
| KNN | 61.42 | 51.31 | 41.65 | 74.28 | 45.98 | 66.17 | 74.28 | 41.65 | 69.99 |
| ANN | 61.54 | 54.30 | 15.40 | 91.57 | 23.99 | 62.45 | 91.57 | 15.40 | 74.26 |

From the above analysis, it can be concluded that the feature extraction of images using HOG features is not very effective, and the highest accuracy is 61.54% using ANN.

*Analysis of GLCM Features*

GLCM is a matrix that represents the grayscale relationship between pixels at each location of the image, either adjacent pixels or pixels at a specified distance pixel. The work finds that among the 14 statistics derived from GLCM, only four statistics (Homogeneity, Correlation, Contrast, and Energy) are uncorrelated, and these four features are easy to compute and give high classification accuracy [53]; Six texture features are studied in detail and concludes that Contrast and Entropy are the most important features [54]. Therefore, in the extraction of GLCM features, this paper calculates four statistical attributes in the grayscale co-occurrence matrix: Homogeneity, Correlation, Contrast, and Entropy.

The main reason for the bad effect of the GLCM features is due to the fact that the image used is a U-Net segmented image. After the image segmented by U-Net, the pixel value of the background region is zero. This significantly affects the image information carried by the grayscale co-occurrence matrix, whose four statistical attributes are very distorted in describing the texture features of the image. The final classification accuracy using GLCM features is 53.57%.

*Analysis of LBP Features*

In the LBP features extraction stage, features are extracted for each U-Net segmentation image using the LBP operator to form a grayscale image with the same resolution as the original image. From the LBP features extracted image, the LBP histogram is formed using the image grayscale, as in Figure 10, the horizontal coordinate is the grayscale of the LBP image, and the vertical coordinate is the number of pixels per grayscale.

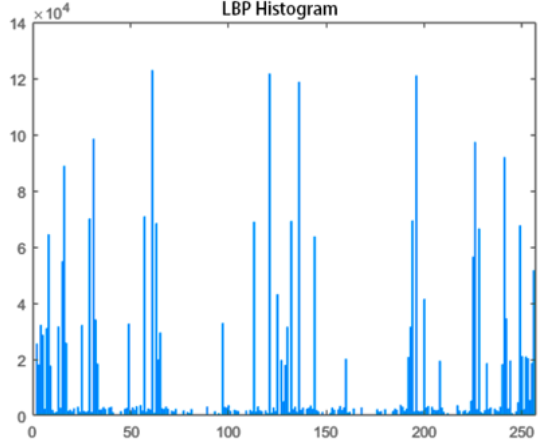

Figure 10: LBP histogram. The horizontal coordinate is the grayscale of the LBP image, and the vertical coordinate is the number of pixels per grayscale.

The advantages of LBP features are fast operation speed and rotation invariance and grayscale invariance. The LBP value of each pixel can reflect the texture relationship between the point and the surrounding pixels, and the texture information is better preserved. The disadvantage is that it is sensitive to the oritation information, the gradient oritation in the image has a relatively large impact on the LBP value of the pixel. The purpose of this study is to classify gastric cancer tissues and normal tissues, and the LBP feature is extracted from the local texture information of the image to form the LBP image. The gray-scale histogram is then drawn from the LBP image. Instead of passing down all the information of the tissue region, the histogram passes down the local texture features, which eventually leads to the loss of the feature information needed for the experiment. The classification accuracy using LBP features is 65.71%.

**Analysis of Classifier Design**

In the classifier design stage, as shown in Table 3, this paper compares five methods. First, the image is segmented by U-Net and then graph based features (MST) are used to extract the features. At last, the classification accuracy of the five classifiers RBF SVM, Linear SVM, ANN, RF and KNN are compared respectively. In this paper, we select a 6-layer ANN network, a RF with 512 trees and a KNN structure with $k$=15.

The classification accuracy in Table 4 shows that the graph based features in this paper have good robustness and the experimentally used classifiers are effective. By comparison, the most suitable for the structural features of the histopathological image of gastric cancer is the RBF SVM classifier.

Table 4: Design method of classifier. The first column shows the serial number. The second column shows the segmentation using U-net. The third column shows the five feature extraction methods used in this paper. The fourth column shows the classification with five different classifiers. The fifth column shows the accuracy rates corresponding to the results of different classifiers in the experiments.

| No. | Segmentation Method | Feature Extraction | Classifier | Accuracy |
|---|---|---|---|---|
| 1 | U-Net Segmentation | (Graph Based Features) $k$-means extraction of cell nuclei MST drawing Extracting edge 8 features | RBF SVM | 94.29% |
| 2 | | | Linear SVM | 75.00% |
| 3 | | | ANN | 90.71% |
| 4 | | | RF | 87.14% |
| 5 | | | KNN | 92.86% |

Meanwhile, a comparison experiment of two types of deep learning classification is designed: VGG-16 and Inception-V3. As shown in Table 5, in order to compare the effect of different deep learning classification methods under the same experimental data, the data of the comparison experiment are 70 randomly selected from each of 140 gastric cancer images and 140 normal images, with a final training set and test set of 140 images each. To better compare the classifier performance, we achieve data augmentation by meshing the current images into patches (256×256 pixels). After augmentation, 8960 training images and the same number of test set are obtained.

Table 5: Comparative experiments in deep learning. The first row of the table shows the serial number, the image segmentation method, the feature extraction method, the classifier and the accuracy rate. The second row shows the optimal processing methods obtained in the above experiments and their accuracy rates. The third row shows the deep learning classification using VGG-16 and its accuracy. The fourth row shows the deep learning classification using Inception-V3 and its accuracy. The fifth row shows the accuracy after the augmentation.

| No. | Segmentation Method | Feature Extraction | Classifier | Accuracy | Accuracy After Augmentation |
|---|---|---|---|---|---|
| 1 | U-Net | Graph Based Features (MST) | RBF SVM | 94.29% | -- |
| 2 | -- | -- | VGG-16 | 75.00% | 87.50% |
| 3 | -- | -- | Inception-V3 | 50.00% | 62.80% |

The advantage of VGG-16 is that the framework is more concise, and the network uses the same volume of convolutional kernels and max-pooling, and a combination of several small filter (3×3) convolutional layers instead of one large filter (5×5 or 7×7). The advantage of Inception-V3 is that it uses different sized convolutional kernels within a layer, which improves perceptual power, and uses batch normalisation, which mitigates gradient disappearance. These two deep learning networks are chosen for the classification comparison experiments in this experiment and after training. But it is found that the classification results are not good, with VGG-16 a classification accuracy of 75% and Inception-V3 a classification accuracy of 50%. The main reason for this is that deep learning networks require a large amount of sample data. After augmentation, the accuracy of both VGG-16 and Inception-V3 network models have been improved, which are 87.50% and 62.80% respectively, but they are still lower than our proposed method.

## DISCUSSION

For the different methods used for feature extraction, this paper compares the different classification results of graph based features with other features and draws separate confusion matrices to analyse them. The main evaluation metrics consist of the following five components: Accuracy (ACC), Precision (PPV), Recall (TPR), Specificity (TNR) and F1-Score. And evaluation metrics are calculated for each confusion matrix, the results of which are shown in Table 6.

Table 6: Evaluation Index (Unit: \%). The first column of the table shows the serial number, the second column shows the five different feature extraction methods, and the third to seventh columns show the values of the evaluation indicators corresponding to each feature extraction method.

| No. | Characterisation Method | Accuracy (ACC) | Precision (PPV) | Recall (TPR) | Specificity (TNR) | F1-Score |
|---|---|---|---|---|---|---|
| 1 | Graph Based Features | 94.29 | 100 | 88.57 | 100 | 93.94 |
| 2 | RGB Features | 69.29 | 76.47 | 55.71 | 82.86 | 64.46 |
| 3 | HOG Features | 55.00 | 58.14 | 35.71 | 74.29 | 44.24 |

| 4 | GLCM Features | 53.57 | 55.81 | 34.29 | 72.86 | 42.48 |
| 5 | LBP Features | 65.71 | 68.97 | 57.14 | 74.29 | 62.50 |

From the above confusion matrix and its evaluation index, we can see that in terms of classification accuracy, the graph based features perform very well, with an accuracy rate of 94.29%. The other features methods are less effective, with RGB features and LBP features at 69.29% and 65.57% respectively, and the worst are HOG features and GLCM features, which are even as low as 55% and 53.57%. In terms of accuracy, the graph based features still perform the optimal, having reached 100%, followed by the RGB features at 76.47%. The accuracy of the LBP features is 68.97%, while the worst are the HOG features and GLCM features at 58.14% and 55.81%. In terms of recall (i.e., classification accuracy of normal images), the value of graph based features is 88.57%, with 8 normal images being misclassified as gastric cancer images. The values of RGB features and LBP features are 55.71% and 57.14%, and the worst are still the HOG features and GLCM features at 35.71% and 34.29% respectively. In terms of specificity (accuracy of classification of gastric cancer images), the graph based features work well and reach 100%, which means that no gastric cancer images are classified wrongly. This is followed by the RGB features at 82.86%. The value of HOG features is equal to the LBP features at 74.29%, and the GLCM features only reach 72.86%. In the part of F-score, the graph based features method is still the optimal performer at 93.94%, followed by the RGB and LBP features at 64.46% and 62.50%, and the worst are still the HOG and GLCM features at 44.24% and 42.48%.

Through the above analysis and discussion, it can be concluded that the graph based features extraction method performs the optimal throughout the experiment, followed by RGB features and LBP features, and the lowest are HOG features and GLCM features. Meanwhile, by analysing the results of the comparison experiments between the classification of normal images and the classification of gastric cancer images, it can be found that the method of graph based features is poor in classifying normal images and easily misclassified normal images into gastric cancer images. But it performs well in classifying gastric cancer images.

## CONCLUSION

Histopathological image analysis has been a popular research direction in the medical field and plays a crucial role in the future path of intelligent medicine. In the study of the topology of histopathological images of gastric cancer, graph theory is able to address the problems in this direction. The histopathological images of gastric cancer have a wide range of tissue structures and complex morphology, especially those in the cancer nest region. And it is difficult to extract the complete tissue information with conventional features to meet the experimental requirements.

In this paper, a graph based features microscopic image analysis method is proposed for gastric cancer histopathology, which expands on the classical digital image processing process and mainly includes the main steps of image segmentation, feature extraction and classifier design. This analysis method mainly takes advantage of the features that the topological structure information of gastric cancer tissue regions is significantly different from normal tissues and uses graph based features to collect this information and then classify it. By comparing the classification result metrics, this paper again validates the advantages of graph based features on histopathological images of gastric cancer. Also, by comparing multiple image segmentation methods, multiple feature extraction methods, multiple classifiers and experiments for deep learning, the optimal method can be selected: for histopathological images of gastric cancer, the image is first segmented using U-net, extracted by the graph based features method and finally the RBF SVM classifier, which is optimal for non-linear data processing, is selected for classification. The final experimental data shows that our analysis method has an absolute advantage in classifying histopathological images of gastric cancer. Furthermore, the proposed graph based features have a potential to work in other microscopic image analysis field, such as microorganism image analysis [55-57], cytopathological image analysis [58-61], microscopic video analysis [62-66].

## Data Availability

The data used to support the findings of this study are available from the corresponding author upon request.

## Conflicts of Interest

The authors declare that they have no conflicts of interest.

## Acknowledgements

This work is supported by National Natural Science Foundation of China (No. 61806047). We thank Miss Zixian Li and Mr. Guoxian Li for their important discussion. We also thank B.E. Jiawei Zhang for his help in the experiments.